\documentclass[twocolumn,aps,pra,amsfonts,amssymb]{revtex4-1}
\usepackage{amsmath}
\usepackage{graphicx,epstopdf}
\usepackage{bm}
\usepackage{array}
\usepackage{dcolumn}
\usepackage{hepparticles}
\usepackage{heppennames}
\usepackage{color}

\newcounter{gg}

\newcommand{\zhat}{\bf \hat{z}}

\newcommand{\pbar}{$\overline{\rm{p}}$~}

\begin{document}

\title{Resolving an Individual One-Proton Spin Flip to Determine a Proton Spin State}

\author{J.\ DiSciacca}
\affiliation{Department of Physics, Harvard University, Cambridge, MA 02138}

\author{M.\ Marshall}
\affiliation{Department of Physics, Harvard University, Cambridge, MA 02138}

\author{K.\ Marable}
\affiliation{Department of Physics, Harvard University, Cambridge, MA 02138}

\author{G.\ Gabrielse}
\email[Email: ]{gabrielse@physics.harvard.edu} \affiliation{Department of Physics, Harvard University, Cambridge, MA
02138}

\date{Accepted by Phys.\ Rev.\ Lett.\ on 7 March 2013}
\begin{abstract}     
Previous measurements with a single trapped proton (p) or antiproton (\pbar) detected spin resonance from the increased scatter of frequency measurements caused by many spin flips.  Here a measured correlation confirms that individual spin transitions and states are detected instead. The high fidelity  suggests that it may be possible to use quantum jump spectroscopy to measure the p and \pbar magnetic moments much more precisely.  
\end{abstract}


\maketitle

\newcommand{\w}{3.00in}
\newcommand{\comment}[1]{\uppercase{\bf #1}}

The fundamental reason for the striking imbalance of matter and antimatter in the universe has yet to be discovered.  Precise comparisons of antimatter and matter particles are thus of interest.  Within the standard model of particle physics, a CPT theorem \cite{Luders57} predicts the relative properties of particles and antiparticles. (The initials represent charge conjugation, parity and time reversal symmetry transformations).  The theorem pertains because systems are described using local, Lorentz-invariant quantum field theory (QFT). Whether the CPT theorem is universal, of course, is open to question since gravity so far eludes a QFT description. A testable prediction is that particles and antiparticles have magnetic moments of the same magnitude and opposite sign.  The moment of a single trapped \pbar \cite{PbarMagneticMoment} was recently measured to a precision 680 times higher than had been possible with other methods.  The ratio of \pbar and p moments is consistent with the CPT prediction to 4.4 ppm.

Quantum jump spectroscopy of a single trapped electron shows that a magnetic moment can be measured much more precisely, to $3$ parts in $10^{13}$ \cite{HarvardMagneticMoment2008}. Individual spin transitions were resolved to determine the needed spin precession frequency.  For the substantially smaller nuclear moments of the \pbar and p this is much more difficult.  This Letter reports the first observation of individual spin transitions and states for a single p in a Penning trap, with a method applicable for a \pbar.  A high 96\% fidelity is realized by selecting a low energy cyclotron motion from a thermal distribution, by saturating the spin transition, and by careful radiofrequency shielding.  The modest spin state detection efficiency realized in this initial demonstration could be used to make a magnetic moment measurement.  However, it now seems possible to use adiabtic passage to detect the spin state in every detection attempt to decrease the measurement time. The possibility to measure a \pbar cyclotron frequency (the other frequency needed to determine the moment) has been demonstrated to $1$ part in $10^{10}$ \cite{FinalPbarMass} to compare the charge-to-mass ratios of the \pbar and p \cite{FinalPbarMass}.  With the spin method demonstrated here, it may be possible to approach this precision in comparing the \pbar and p magnetic moments to make a second precise test of the CPT theorem with a baryon.

\begin{figure}[htbp!]
\centering
\includegraphics*[width=\columnwidth]{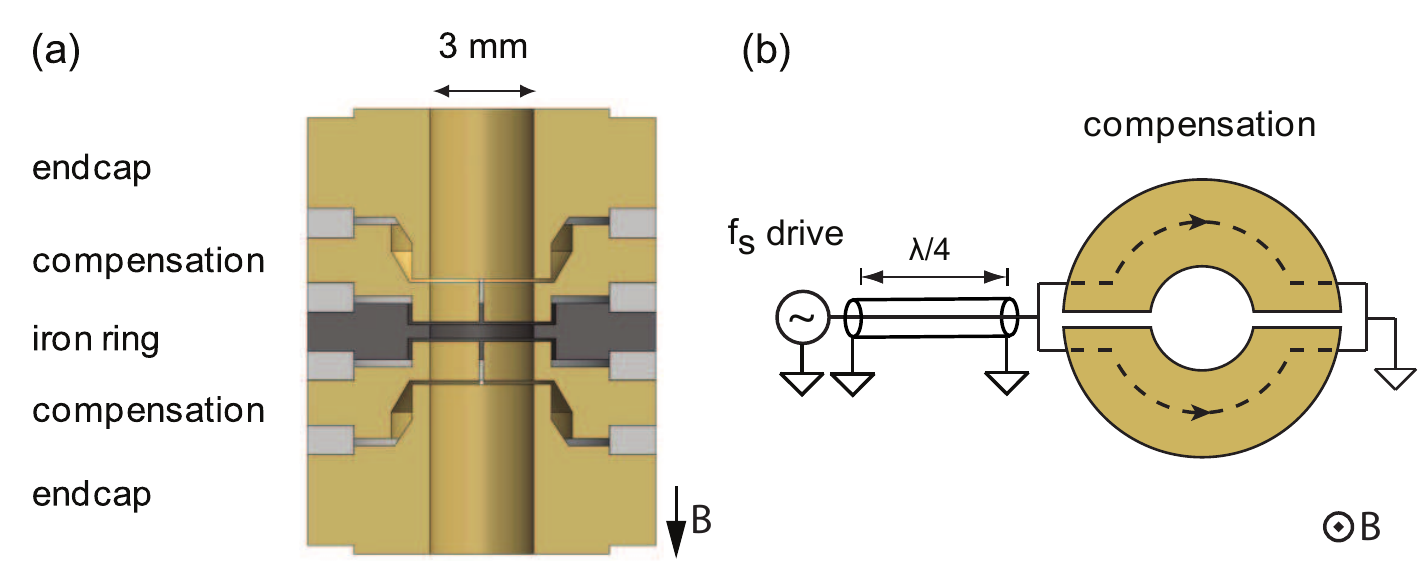}
\caption{(a) Cutaway side view of Penning trap electrodes.  All are copper except for an iron ring that makes the magnetic gradient needed to observe a spin flip. (b) Top view of the oscillating current paths for the spin flip drive.
}
\label{fig:Trap}
\end{figure}

The trap electrodes in Fig.~\ref{fig:Trap} have already been used with both a p and a \pbar.  They were used in 2011 to measure the p magnetic moment \cite{ProtonMagneticMoment}, in early 2012 for this p demonstration, and then in mid 2012 were moved to CERN to measure the \pbar magnetic moment \cite{PbarMagneticMoment}. Leaving details to the other reports, the p is suspended at the center of an iron ring electrode sandwiched between OFE copper electrodes.  The electrodes have gold evaporated on their surfaces.  Thermal contact with liquid helium keeps them at 4.2 K and gives a vacuum that essentially eliminates collisions with background gas atoms.  Voltages applied to electrodes with a carefully chosen relative geometry \cite{OpenTrap} give a high quality electrostatic quadrupole potential while allowing the proton to be moved into the trap through the open access from either end.   

In a magnetic field $\mathbf{B} \approx -5 \,\zhat$ Tesla, vertical in Fig.~\ref{fig:Trap}a, the protons's spin up and down energy levels are separated by  $h f_s$, with a spin precession frequency $f_s = 221.35$ MHz. The proton energy in the magnetic field is higher for a spin that is up with respect to the quantization axis $\zhat$ than for a spin down.  A driving force that can flip the spin involves a magnetic field perpendicular to {\bf B} that oscillates at approximately $f_s$.  This field is generated by  currents sent through halves of a compensation electrode (Fig.~\ref{fig:Trap}b).  The trapped proton's circular cyclotron motion is perpendicular to {\bf B} with a frequency $f_+ = 79.26$ MHz
slightly shifted from $f_c$  by the electrostatic potential. The proton also oscillates parallel to {\bf B} at about $f_z =
\,919$ kHz.    The proton's third motion is a circular magnetron motion, also perpendicular to
{\bf B}, at the much lower frequency $f_- = 5.28$ kHz.

Small measured shifts in the axial frequency $f_z$, 
\begin{equation} 
\Delta f_z \propto \left( n + \frac{1}{2} + \frac{g_p m_s}{2} + \frac{f_-}{f_+} [\ell + \frac{1}{2}] \right),  \label{eq:FrequencyShift}
\end{equation}
reveal changes in the cyclotron, spin and magnetron quantum numbers $n$, $m_s$ and $\ell$ \cite{Review}.  The shifts are taken to be the shifts in the self-excited oscillation (SEO) that arises when amplified signal from the proton's axial oscillation is fed back to drive the p into a steady-state oscillation \cite{OneProtonSelfExcitedOscillator}.
The shifts arise as the magnetic moments of these motions interact with a magnetic bottle gradient from the saturated iron ring, 
\begin{equation}
\Delta {\bf B} = \beta_2 [(z^2-\rho^2/2){\bf \hat{z}} - z\rho \boldsymbol{\hat{\rho}}],
\end{equation}
with $\beta_2 = 2.9 \times 10^5$ T/m$^2$. A spin flip causes only a tiny shift, $\Delta_s=130$ mHz, despite the gradient being 
190 times larger than used to detect electron spin flips \cite{HarvardMagneticMoment2008}, because a nuclear moment is smaller than an electron moment by of order 1/2000, the ratio of the electron and proton masses.

\begin{figure}[htbp!]
\includegraphics*[width=1.0\columnwidth]{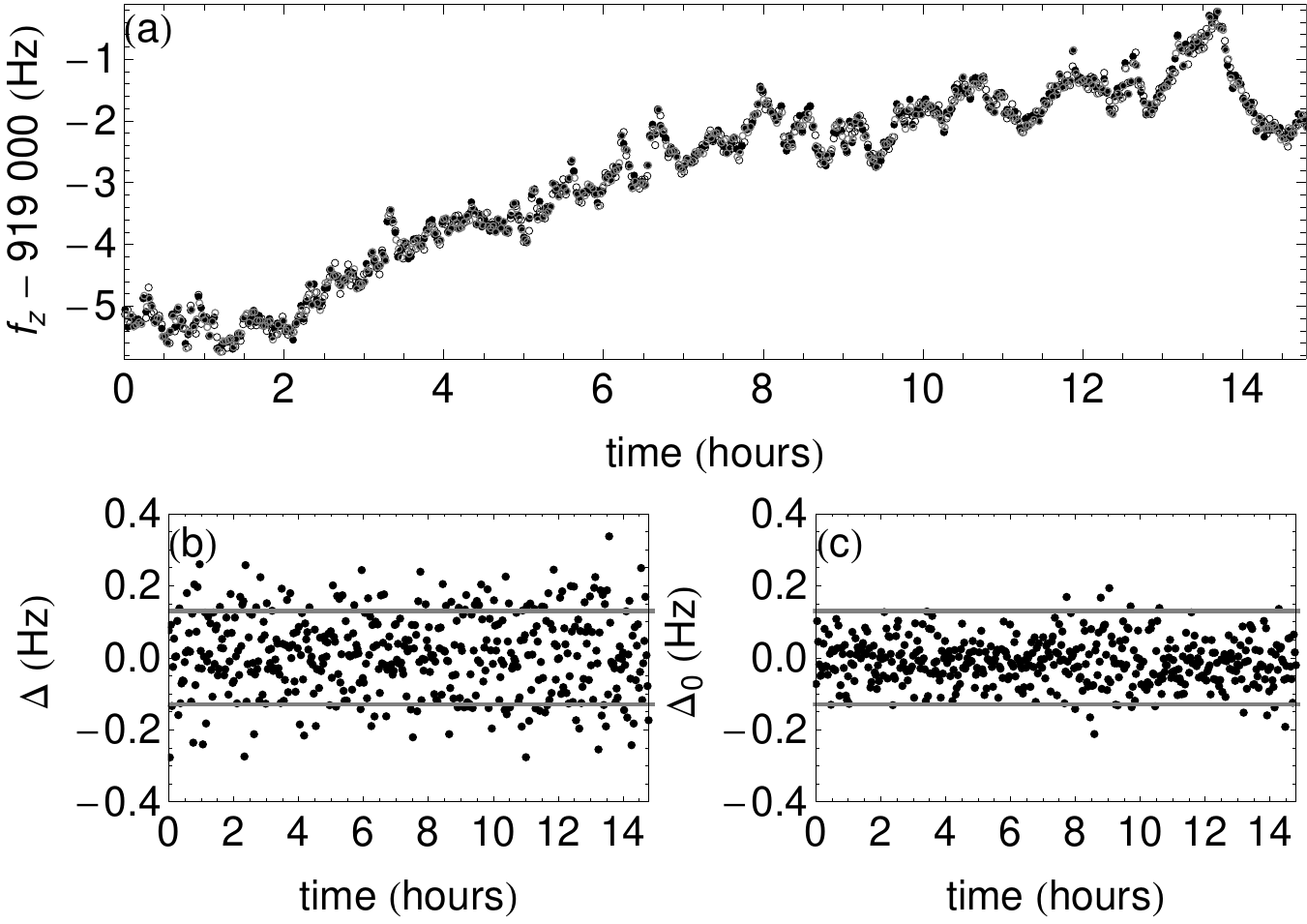}
\caption{(a) Repeated measurements of $f_z$ show a large drift and scatter.  (b) Scatter in the measured frequency shifts $\Delta$ for a resonant spin drive has $\sigma=109$ mHz.  (c) Scatter in the measured frequency shifts $\Delta_0$ for an off-resonant spin drive that causes no spin flips has  $\sigma_0 = 63$ mHz. The gray lines show the spin flip shift $\pm \Delta_s$.} 
\label{fig:FrequencyMeasurements}
\end{figure}

\begin{figure}[htbp!]
\includegraphics*[width=\columnwidth]{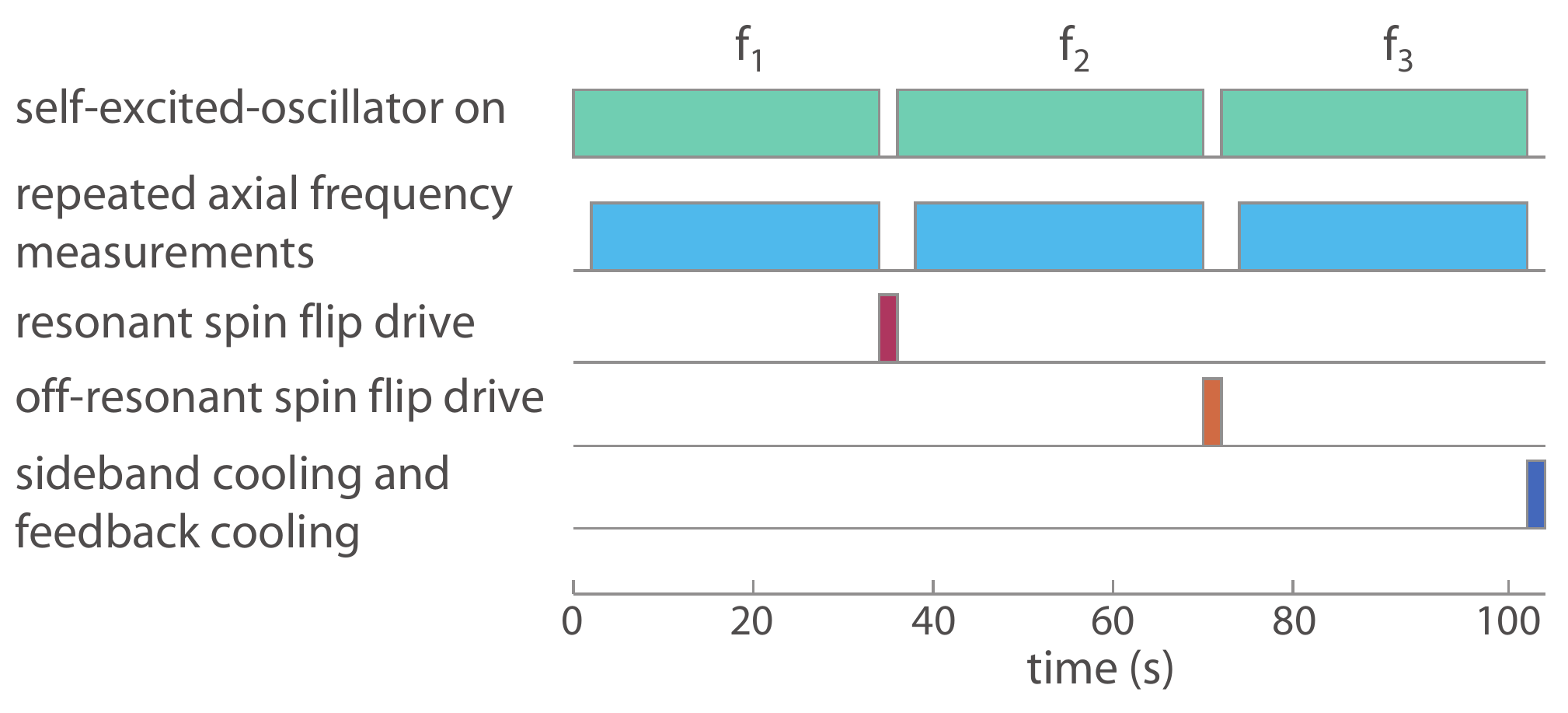}
\caption{Spin detection cycle  repeated nearly 15 hours.}
\label{fig:MeasurementSequence}
\end{figure}

Counting individual spin flips for quantum jump spectroscopy requires identifying the small shifts $\pm \Delta_s$. The nearly 15 hours of $f_z$ measurements in Fig.~\ref{fig:FrequencyMeasurements}a illustrate the challenge of observing such small shifts despite much larger frequency drifts and fluctuations.  Repeated applications of a detection cycle (Fig.~\ref{fig:MeasurementSequence}) yield a series of frequency shifts $\Delta=f_2-f_1$ that take place for a resonant spin drive (Fig.~\ref{fig:FrequencyMeasurements}b) and a series of shifts $\Delta_0 =f_3-f_2$ for a non-resonant spin drive (Fig.~\ref{fig:FrequencyMeasurements}c).  The $f_i$ are averages of the SEO frequency for three 32s periods.  In the 4 s intervals between the averaging periods, the SEO is off and either a resonant or non-resonant (detuned 100 kHz) spin-flip drive is applied for the first 2 s.

The detection cycle concludes with 2 s of sideband cooling and feedback cooling that prevents the average magnetron radius from growing.  Each cooling application however, establishes a slightly different magnetron radius that cannot be predicted \cite{OneProtonSelfExcitedOscillator}, giving here a $137 \pm 5$ mHz spread of $f_z$ values that is comparable to $\Delta_s$.

\begin{figure}[htbp!]
\includegraphics*[width=1.0\columnwidth]{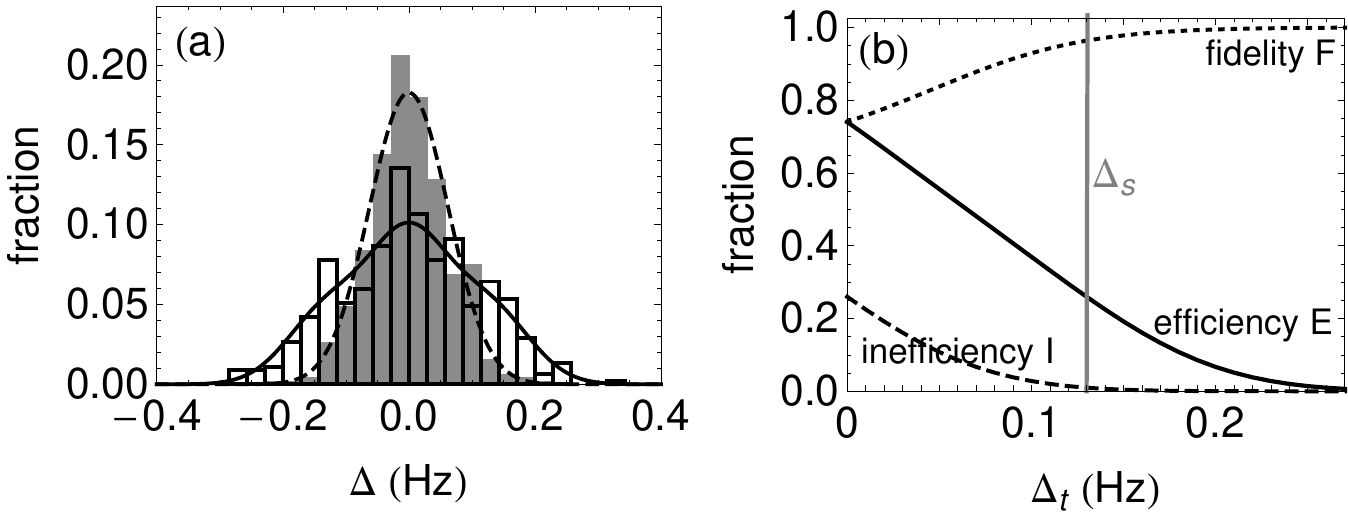}
\caption{(a) The gray histogram of measured changes $\Delta_0$ with no spin flip drive fit well to a Gaussian (dashed).  
The predicted histogram shape for a resonant drive that saturates the spin transition (solid curve), and the measured open histogram.  (b) $E$, $I$ and $F$ for a detection cycle that employs a resonant spin drive that saturates the spin transition.  
}
\label{fig:HistogramsEfficiencyFidelity}
\end{figure}

The distribution of fluctuations $\Delta_0$ observed without spin flips (the gray histogram in Fig.~\ref{fig:HistogramsEfficiencyFidelity}a derived from Fig.~\ref{fig:FrequencyMeasurements}c) fits well to a normalized Gaussian probability function $G(\Delta_0,\sigma_0)$ with a standard deviation $\sigma_0 = 63$ mHz.  This is significantly smaller than the 112 and 145 mHz for the p and \pbar measurements \cite{ProtonMagneticMoment,PbarMagneticMoment}.  (The Allen deviation used in \cite{ProtonMagneticMoment,PbarMagneticMoment} is smaller by $\sqrt{2}$.) Though $\Delta_0$ is larger than we would like, a distribution this narrow requires a p with an unusually small cyclotron orbit, since the fluctuations are observed to increase linearly with cyclotron radius \cite{ProtonMagneticMoment}.  A p is repeatedly transferred between the trap of Fig.~\ref{fig:Trap} and a coaxial trap whose attached circuit damps the cyclotron motion, until a p with a cyclotron energy below the thermal average is selected. Reducing $\sigma_0$ is complicated because the causes of the fluctuations are difficult to identify and control \cite{OneProtonSelfExcitedOscillator}.  One candidate is noise that makes it past considerable radiofrequency shielding to drive the cyclotron motion, with a single quantum change shifting $f_z$ by 50 mHz.    

We can predict the distribution of shifts $\Delta$ for a long series of detection cycles when the resonant spin drive is strong enough to saturate the spin transition.  Half of the detection cycles should produce no spin flip and thus have a distribution of $\Delta$ given by $G(\Delta,\sigma_0)/2$.  A quarter each of the detection cycles should involve spin up and spin down transitions described by $G(\Delta \mp \Delta_s,\sigma_0)/4$, since the spin changes add shifts $\pm\Delta_s$ to the random fluctuations $\Delta_0$ observed when no spin is flipped.  

The sum of the three predicted distributions is the solid curve in Fig.~\ref{fig:HistogramsEfficiencyFidelity}a.  Our interpretation is supported by the good agreement with the open histogram in Fig.~\ref{fig:HistogramsEfficiencyFidelity}a derived from the observed $\Delta$ in Fig.~\ref{fig:FrequencyMeasurements}b.  The observed standard deviation has a $\sigma = 109$ mHz, clearly larger than $\sigma_0$ for the gray histogram for no spin flips.  For the \pbar magnetic moment measurement \cite{PbarMagneticMoment} and related p studies  \cite{ProtonMagneticMoment,MainzSpinFlips,MainzProtonMagneticMoment} the increase from $\sigma_0$ to $\sigma$  is used to find spin resonance with no individual spin flip being resolved. Here, encouraged by the good agreement of the prediction and the observation, we first argue that we are able to identify spin flips from the individual $\Delta$ values in Fig.~\ref{fig:FrequencyMeasurements}b, and then confirm this assertion using a measured spin correlation function.   

Each $\Delta$ would unambiguously reveal which spin flip had occurred, if any, if the $\Delta_0$ for the off-resonance drive fluctuated much less than the spin flip shift $\Delta_s$, so $\sigma_0 \ll \Delta_s$.  In this limit the open histogram would be 3 resolved histograms, each with a width characterized by $\sigma_0$.  The much larger electron magnetic moment makes this possible for measuring the electron moment \cite{HarvardMagneticMoment2008}. 

More care is required for p and \pbar.  Since $\sigma_0=63$ mHz is only half of $\Delta_s=130$ mHz, some fluctuations will be able to hide whether a spin flip shift $\pm\Delta_s$ has taken place.  For a detection cycle that flips the spin state with probability $P$, the 4 ways to produce an above threshold $\Delta \ge \Delta_t$ for positive $\Delta_t>0$  have probabilities  
\begin{eqnarray}
P_{\downarrow \uparrow}(\Delta_t) &=& P \int_{\Delta_t}^{\infty}G(\Delta-\Delta_s,\sigma_0)~d\Delta, \label{eq:du}\\
P_{\uparrow \uparrow}(\Delta_t)=P_{\downarrow \downarrow}(\Delta_t) &=& (1-P) \int_{\Delta_t}^{\infty} G(\Delta,\sigma_0)~d\Delta, \label{eq:uu}\\
P_{\uparrow \downarrow}(\Delta_t) &=& P \int_{\Delta_t}^{\infty} G(\Delta+\Delta_s,\sigma_0)~d\Delta. \label{eq:ud}
\end{eqnarray} 
The largest, $P_{\downarrow \uparrow}(\Delta_t)$, is for a detection cycle that flips the spin from down to up. The probabilities $P_{\downarrow \downarrow}(\Delta_t) = P_{\uparrow \uparrow}(\Delta_t)$ are smaller, and $P_{\uparrow \downarrow}(\Delta_t)$ is smaller still.        

A detection cycle produces an above threshold shift  $\Delta\ge\Delta_t$ with an efficiency $E$ for a spin that is down before the cycle begins, and with an efficiency $I$ for a spin that is instead up before the cycle begins, with   
\begin{eqnarray}
E&=&P_{\downarrow \uparrow}(\Delta_t)+P_{\downarrow \downarrow}(\Delta_t)\\
I &=& P_{\uparrow \uparrow}(\Delta_t) + P_{\uparrow \downarrow}(\Delta_t).
\end{eqnarray} 
The latter is thus an inefficiency with respect to detecting a spin that was initially down.  
The fidelity $F=E/(E+I)$ represents the reliability with which we determine the spin state.  It is the fraction of above threshold events that result from a spin that starts down when the detection cycle is applied.   The same values of $E$, $I$ and $F$ pertain for ``above'' threshold events $\Delta \le-\Delta_t$ observed when a single detection cycle is applied to a spin up.    

The dependence of $E$, $I$ and $F$ upon the choice of threshold $\Delta_t$ is shown in Fig.~\ref{fig:HistogramsEfficiencyFidelity}b for a resonant drive that saturates the spin transition (i.e. $P=1/2$), along with our $\Delta_s$ and $\sigma_0$.  Choosing a threshold equal to the spin flip shift, $\Delta_t = \Delta_s$, gives a high fidelity $F=96$\% and a low $I=1$\%.   However, the efficiency $E=26$\% means that a spin down will produce an above threshold event that establishes the spin state with this high fidelity about in 1 in 4 attempts. Roughly speaking, half of the detection cycles flip the spin as needed to get an above threshold event, and half of these cycles have fluctuations of the same sign as the spin flip shift.   

\begin{figure}[htbp!]
\includegraphics*[width=1.0\columnwidth]{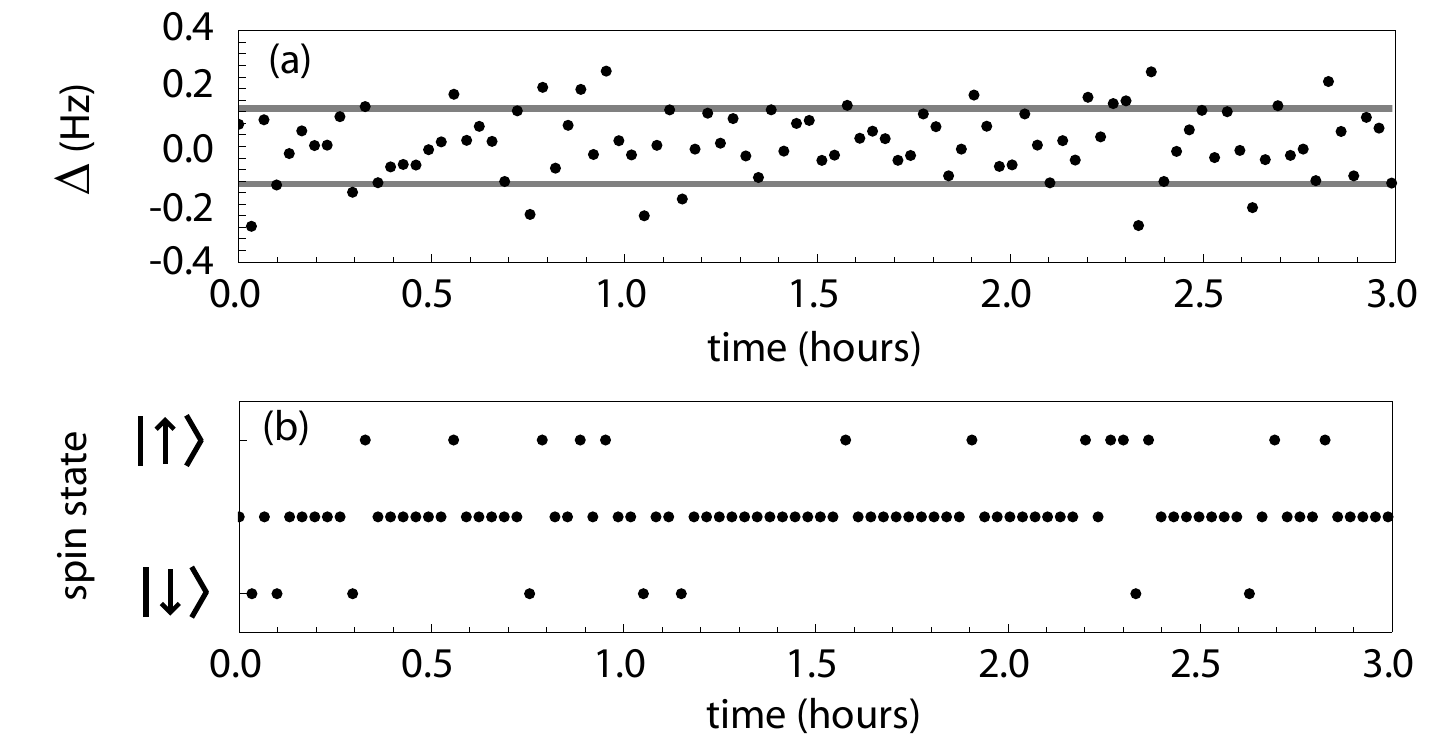}
\caption{(a) Three hour sample of frequency shifts $\Delta$ (from Fig.~\ref{fig:FrequencyMeasurements}b).  (b) Corresponding identifications of the spin state based upon above-threshold $\Delta$ for a threshold of $\Delta_t=\Delta_s$.  Points between the heights of the identified spin states indicate that no spin state identification could be made with this threshold choice.}
\label{fig:SpinFlips}
\end{figure}

A three hour slice of $\Delta$ measurements (from Fig.~\ref{fig:FrequencyMeasurements}b) is shown in Fig.~\ref{fig:SpinFlips}a.  Below, in Fig.~\ref{fig:SpinFlips}b, are spin state determinations (at the end of the detection cycles) made using a threshold $\Delta_t=\Delta_s$ to get a fidelity of 96\% for about 1 in 4 detection cycles.

\begin{figure}[htbp!]
\includegraphics*[width=1.0\columnwidth]{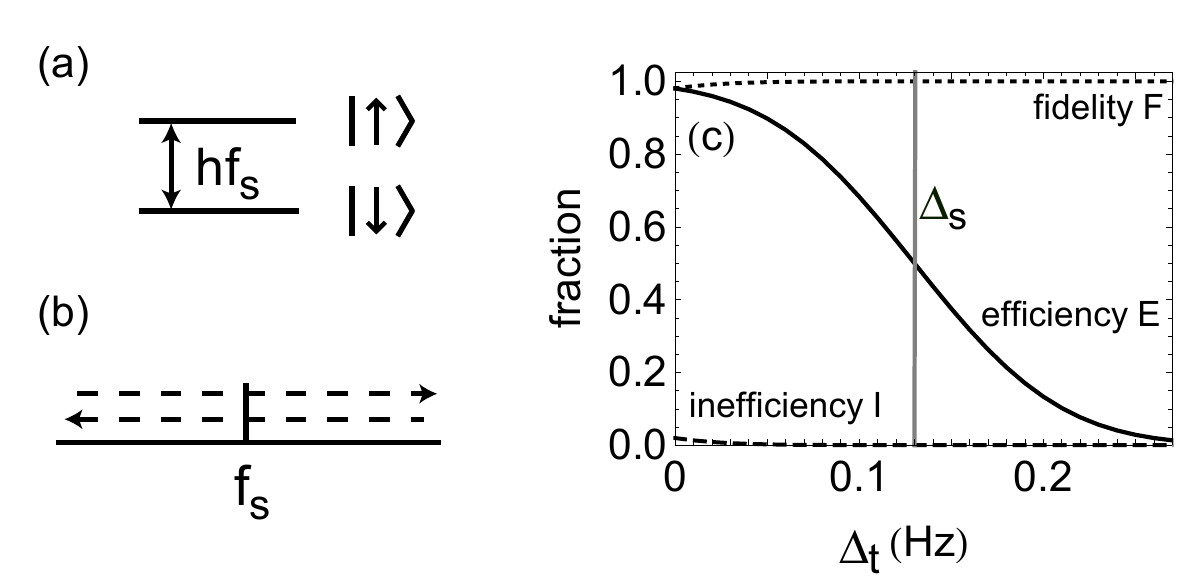}
\caption{(a) Spin energy levels. (b) For adiabatic passage the drive frequency is swept adiabatically upward or downward through resonance.  (c) The efficiency $E$, inefficiency $I$ and fidelity $F$ for an adiabatic passage spin drive applied during the detection cycle.}
\label{fig:AdiabaticPassage}
\end{figure}

A nearly perfect detection efficiency (with the spin state determined in each detection cycle rather than in 1 of 4 for this simple first demonstration) should be possible with an enhanced detection cycle.
We propose to substitute an adiabatic passage drive (or a less robust $\pi$ pulse) for the simple resonant drive to increase the spin flip probability from $P=1/2$ to  $P=1$.  No reduction in $\sigma_0=63$ mHz is required. As demonstrated decades ago in NMR measurements, complete population transfer from one state to the other in Fig.~\ref{fig:AdiabaticPassage}a can be accomplished by sweeping the drive adiabatically either upwards or downwards through resonance (Fig.~\ref{fig:AdiabaticPassage}b). Fig.~\ref{fig:AdiabaticPassage}c shows how the fidelity and efficiency depend on threshold.  A threshold of $\Delta_t=0$ mHz, for example, gives a nearly perfect fidelity $F=98\%$ and efficiency $E=98\%$.   The care that must be taken to minimize the possible disruption of population transfer from thermal axial motion in the magnetic gradient is under study.

\begin{figure}[htbp!]
\includegraphics*[width=1.0\columnwidth]{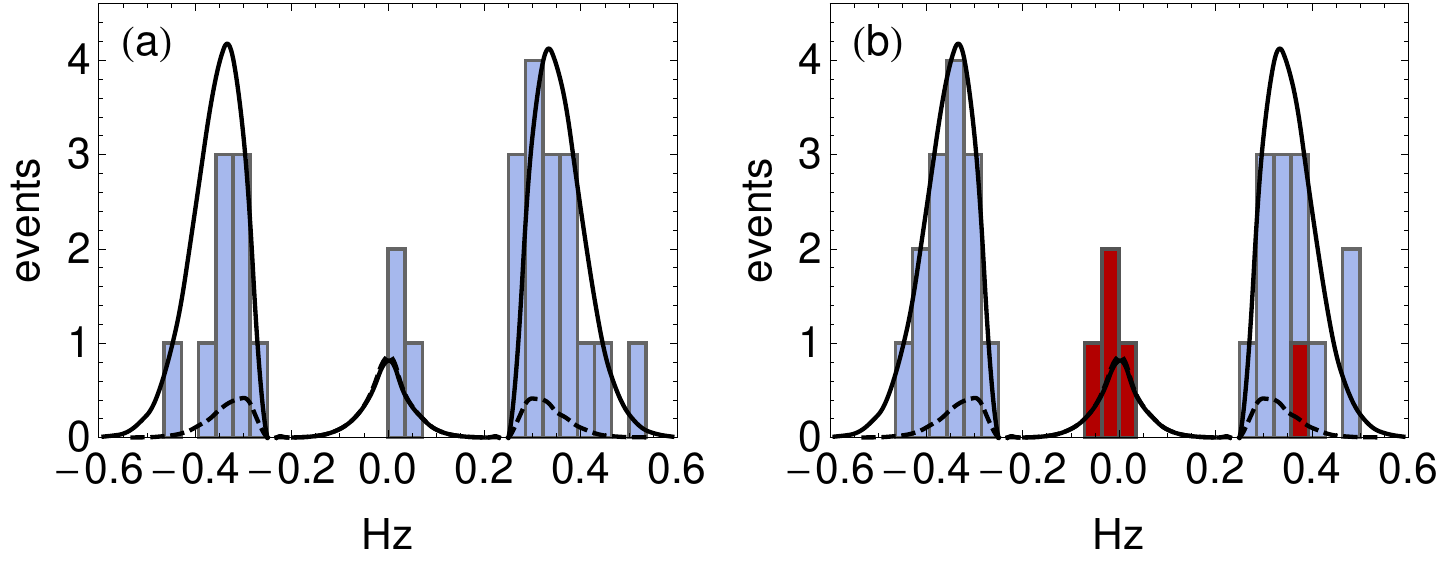}
\caption{The solid curve is the predicted shape of the correlation histogram for adjacent above-threshold events, and the dashed curve is the predicted distribution of accidentals. The histogram for the 450 detection cycles in our data set (a) agrees qualitatively and quantitatively with our predictions.  The histogram in (b) is a simulation for 450 detection cycles.  The 5 accidentals are highlighted.}
\label{fig:CorrelationFunctions}
\end{figure}

Confirming evidence that individual spin flips are being observed comes from a measured correlation function  (Fig.~\ref{fig:CorrelationFunctions}a) that is qualitatively and quantitatively consistent with predictions.  We use correlations $\Delta_2-\Delta_1$ that come from a detection cycle that produces an above-threshold $\Delta_1$, followed immediately by a second detection cycle that also produces an above-threshold $\Delta_2$.    For the 450 detection cycles of our data set, with the observed $\sigma_0=63$ mHz and chosen threshold $\Delta_t=\Delta_s$, there are about $E\,450\approx 120$ above-threshold $\Delta$ (with either $\Delta\ge\Delta_t$ or $\Delta\le -\Delta_t$).  About $E^2\,450\approx 30$ pairs of these are produced by sequential detection cycles and thus contribute to Fig.~\ref{fig:CorrelationFunctions}a. 

Qualitatively, a histogram of these correlations should have half of its entries below $-2\Delta_s$ (for a spin that flips from up to down in the first cycle and from down to up in the next).  The other half of the entries should be in a peak above $2\Delta_s$ (for a spin that flips from down to up in the first cycle and from up to down in the next).  Ideally there should be no entries between the peaks since correlations near zero would require a spin to switch from either up to down or down to up in both of the cycles and this is not possible. However, because the the fidelity is not perfect, some accidentals are expected between the peaks and elsewhere.  These are entries for which one or both of the above-threshold events is due to unusually large fluctuations rather than from a spin flip.

Quantitative predictions come from simulations.  The solid curve in Fig.~\ref{fig:CorrelationFunctions} gives the predicted shapes of the correlation histogram for the measured $\sigma_0$ and a threshold choice $\Delta_t=\Delta_s$.  The dashed curve, the predicted distribution of accidentals, shows that the small central peak is entirely from accidentals since for this peak the solid and dashed curves overlap. 

The measured correlation histogram in Fig.~\ref{fig:CorrelationFunctions}a for the 450 detection cycles of our data set agrees well with the prediction.  It  has $25$ counts in the side peaks and 3 in the center, consistent with the predicted $30\pm 7$ in the side peaks (with $2\pm 2$ of these from accidentals) along with $2 \pm 2$ in the central peak from accidentals.  

Fig.~\ref{fig:CorrelationFunctions}b shows one of many simulated correlation histograms for 450 detection cycles, with 5 accidentals highlighted to distinguish them.  From many such trials we get the mean number and uncertainty for the number of counts in each peak and for the number of accidentals.  

In conclusion, the correlation histogram adds convincing evidence that individual proton spin flips are being observed and well understood.  Individual spin flips of a single trapped proton are observed as above-threshold frequency shifts produced  using a detection cycle that employs the simplest saturated spin drive.  The 96\% fidelity achieved in this first demonstration makes it possible to identify the spin state for 1 in 4 detection cycles.  A substantial increase in this efficiency is predicted when an adiabatic passage drive is substituted for the resonant drive in the detection cycle.  The observations of single proton spin flips open the possibility of quantum jump spectroscopy measurement of the spin frequency for a \pbar or p, to go with precise measurements of their cyclotron frequency demonstrated earlier.  It may eventually be possible to measure these frequencies precisely enough to determine the proton and antiproton magnetic moments  a factor of $10^3$ to $10^4$ times more precisely than achieved in the recent measurement of the \pbar magnetic moment -- itself a     680-fold improvement in precision compared to previous measurements.

Thanks to the NSF AMO program and the AFOSR for support, and to S.\ Ettenauer and E.\ Tardiff for helpful comments on the manuscript.


\begin{thebibliography}{10}%
\makeatletter
\providecommand \@ifxundefined [1]{%
 \@ifx{#1\undefined}
}%
\providecommand \@ifnum [1]{%
 \ifnum #1\expandafter \@firstoftwo
 \else \expandafter \@secondoftwo
 \fi
}%
\providecommand \@ifx [1]{%
 \ifx #1\expandafter \@firstoftwo
 \else \expandafter \@secondoftwo
 \fi
}%
\providecommand \natexlab [1]{#1}%
\providecommand \enquote  [1]{``#1''}%
\providecommand \bibnamefont  [1]{#1}%
\providecommand \bibfnamefont [1]{#1}%
\providecommand \citenamefont [1]{#1}%
\providecommand \href@noop [0]{\@secondoftwo}%
\providecommand \href [0]{\begingroup \@sanitize@url \@href}%
\providecommand \@href[1]{\@@startlink{#1}\@@href}%
\providecommand \@@href[1]{\endgroup#1\@@endlink}%
\providecommand \@sanitize@url [0]{\catcode `\\12\catcode `\$12\catcode
  `\&12\catcode `\#12\catcode `\^12\catcode `\_12\catcode `\%12\relax}%
\providecommand \@@startlink[1]{}%
\providecommand \@@endlink[0]{}%
\providecommand \url  [0]{\begingroup\@sanitize@url \@url }%
\providecommand \@url [1]{\endgroup\@href {#1}{\urlprefix }}%
\providecommand \urlprefix  [0]{URL }%
\providecommand \Eprint [0]{\href }%
\providecommand \doibase [0]{http://dx.doi.org/}%
\providecommand \selectlanguage [0]{\@gobble}%
\providecommand \bibinfo  [0]{\@secondoftwo}%
\providecommand \bibfield  [0]{\@secondoftwo}%
\providecommand \translation [1]{[#1]}%
\providecommand \BibitemOpen [0]{}%
\providecommand \bibitemStop [0]{}%
\providecommand \bibitemNoStop [0]{.\EOS\space}%
\providecommand \EOS [0]{\spacefactor3000\relax}%
\providecommand \BibitemShut  [1]{\csname bibitem#1\endcsname}%
\let\auto@bib@innerbib\@empty
\bibitem [{\citenamefont {L{\"u}ders}(1957)}]{Luders57}%
  \BibitemOpen
  \bibfield  {author} {\bibinfo {author} {\bibfnamefont {G.}~\bibnamefont
  {L{\"u}ders}},\ }\href@noop {} {\bibfield  {journal} {\bibinfo  {journal}
  {Ann. Phys.}\ }\textbf {\bibinfo {volume} {2}},\ \bibinfo {pages} {1}
  (\bibinfo {year} {1957})}\BibitemShut {NoStop}%
\bibitem [{\citenamefont {DiSciacca}\ and\ \citenamefont {{et
  al.}}(2013)}]{PbarMagneticMoment}%
  \BibitemOpen
  \bibfield  {author} {\bibinfo {author} {\bibfnamefont {J.}~\bibnamefont
  {DiSciacca}}\ and\ \bibinfo {author} {\bibnamefont {{et al.}}},\ }\href@noop
  {} {\bibfield  {journal} {\bibinfo  {journal} {Phys. Rev. Lett.}\ }\textbf
  {\bibinfo {volume} {xxx}},\ \bibinfo {pages} {xxx} (\bibinfo {year}
  {2013})}\BibitemShut {NoStop}%
\bibitem [{\citenamefont {Hanneke}\ \emph {et~al.}(2008)\citenamefont
  {Hanneke}, \citenamefont {Fogwell},\ and\ \citenamefont
  {Gabrielse}}]{HarvardMagneticMoment2008}%
  \BibitemOpen
  \bibfield  {author} {\bibinfo {author} {\bibfnamefont {D.}~\bibnamefont
  {Hanneke}}, \bibinfo {author} {\bibfnamefont {S.}~\bibnamefont {Fogwell}}, \
  and\ \bibinfo {author} {\bibfnamefont {G.}~\bibnamefont {Gabrielse}},\
  }\href@noop {} {\bibfield  {journal} {\bibinfo  {journal} {Phys. Rev. Lett.}\
  }\textbf {\bibinfo {volume} {100}},\ \bibinfo {pages} {120801} (\bibinfo
  {year} {2008})}\BibitemShut {NoStop}%
\bibitem [{\citenamefont {Gabrielse}\ \emph {et~al.}(1999)\citenamefont
  {Gabrielse}, \citenamefont {Khabbaz}, \citenamefont {Hall}, \citenamefont
  {Heimann}, \citenamefont {Kalinowsky},\ and\ \citenamefont
  {Jhe}}]{FinalPbarMass}%
  \BibitemOpen
  \bibfield  {author} {\bibinfo {author} {\bibfnamefont {G.}~\bibnamefont
  {Gabrielse}}, \bibinfo {author} {\bibfnamefont {A.}~\bibnamefont {Khabbaz}},
  \bibinfo {author} {\bibfnamefont {D.~S.}\ \bibnamefont {Hall}}, \bibinfo
  {author} {\bibfnamefont {C.}~\bibnamefont {Heimann}}, \bibinfo {author}
  {\bibfnamefont {H.}~\bibnamefont {Kalinowsky}}, \ and\ \bibinfo {author}
  {\bibfnamefont {W.}~\bibnamefont {Jhe}},\ }\href@noop {} {\bibfield
  {journal} {\bibinfo  {journal} {Phys. Rev. Lett.}\ }\textbf {\bibinfo
  {volume} {82}},\ \bibinfo {pages} {3198} (\bibinfo {year}
  {1999})}\BibitemShut {NoStop}%
\bibitem [{\citenamefont {DiSciacca}\ and\ \citenamefont
  {Gabrielse}(2012)}]{ProtonMagneticMoment}%
  \BibitemOpen
  \bibfield  {author} {\bibinfo {author} {\bibfnamefont {J.}~\bibnamefont
  {DiSciacca}}\ and\ \bibinfo {author} {\bibfnamefont {G.}~\bibnamefont
  {Gabrielse}},\ }\href@noop {} {\bibfield  {journal} {\bibinfo  {journal}
  {Phys. Rev. Lett.}\ }\textbf {\bibinfo {volume} {108}},\ \bibinfo {pages}
  {153001} (\bibinfo {year} {2012})}\BibitemShut {NoStop}%
\bibitem [{\citenamefont {Gabrielse}\ \emph {et~al.}(1989)\citenamefont
  {Gabrielse}, \citenamefont {Haarsma},\ and\ \citenamefont
  {Rolston}}]{OpenTrap}%
  \BibitemOpen
  \bibfield  {author} {\bibinfo {author} {\bibfnamefont {G.}~\bibnamefont
  {Gabrielse}}, \bibinfo {author} {\bibfnamefont {L.}~\bibnamefont {Haarsma}},
  \ and\ \bibinfo {author} {\bibfnamefont {S.~L.}\ \bibnamefont {Rolston}},\
  }\href@noop {} {\bibfield  {journal} {\bibinfo  {journal} {Intl. J. of Mass
  Spec. and Ion Proc.}\ }\textbf {\bibinfo {volume} {88}},\ \bibinfo {pages}
  {319} (\bibinfo {year} {1989})},\ \bibinfo {note} {ibid. {93:}, 121
  (1989)}\BibitemShut {NoStop}%
\bibitem [{\citenamefont {Brown}\ and\ \citenamefont
  {Gabrielse}(1986)}]{Review}%
  \BibitemOpen
  \bibfield  {author} {\bibinfo {author} {\bibfnamefont {L.~S.}\ \bibnamefont
  {Brown}}\ and\ \bibinfo {author} {\bibfnamefont {G.}~\bibnamefont
  {Gabrielse}},\ }\href@noop {} {\bibfield  {journal} {\bibinfo  {journal}
  {Rev. Mod. Phys.}\ }\textbf {\bibinfo {volume} {58}},\ \bibinfo {pages} {233}
  (\bibinfo {year} {1986})}\BibitemShut {NoStop}%
\bibitem [{\citenamefont {Guise}\ \emph {et~al.}(2010)\citenamefont {Guise},
  \citenamefont {DiSciacca},\ and\ \citenamefont
  {Gabrielse}}]{OneProtonSelfExcitedOscillator}%
  \BibitemOpen
  \bibfield  {author} {\bibinfo {author} {\bibfnamefont {N.}~\bibnamefont
  {Guise}}, \bibinfo {author} {\bibfnamefont {J.}~\bibnamefont {DiSciacca}}, \
  and\ \bibinfo {author} {\bibfnamefont {G.}~\bibnamefont {Gabrielse}},\
  }\href@noop {} {\bibfield  {journal} {\bibinfo  {journal} {Phys. Rev. Lett.}\
  }\textbf {\bibinfo {volume} {104}},\ \bibinfo {pages} {143001} (\bibinfo
  {year} {2010})}\BibitemShut {NoStop}%
\bibitem [{\citenamefont {Ulmer}\ \emph {et~al.}(2011)\citenamefont {Ulmer},
  \citenamefont {Rodegheri}, \citenamefont {Blaum}, \citenamefont {Kracke},
  \citenamefont {Mooser}, \citenamefont {Quint},\ and\ \citenamefont
  {Walz}}]{MainzSpinFlips}%
  \BibitemOpen
  \bibfield  {author} {\bibinfo {author} {\bibfnamefont {S.}~\bibnamefont
  {Ulmer}}, \bibinfo {author} {\bibfnamefont {C.~C.}\ \bibnamefont
  {Rodegheri}}, \bibinfo {author} {\bibfnamefont {K.}~\bibnamefont {Blaum}},
  \bibinfo {author} {\bibfnamefont {H.}~\bibnamefont {Kracke}}, \bibinfo
  {author} {\bibfnamefont {A.}~\bibnamefont {Mooser}}, \bibinfo {author}
  {\bibfnamefont {W.}~\bibnamefont {Quint}}, \ and\ \bibinfo {author}
  {\bibfnamefont {J.}~\bibnamefont {Walz}},\ }\href@noop {} {\bibfield
  {journal} {\bibinfo  {journal} {Phys. Rev. Lett.}\ }\textbf {\bibinfo
  {volume} {106}},\ \bibinfo {pages} {253001} (\bibinfo {year}
  {2011})}\BibitemShut {NoStop}%
\bibitem [{\citenamefont {Rodegheri}\ \emph {et~al.}(2012)\citenamefont
  {Rodegheri}, \citenamefont {Blaum}, \citenamefont {Kracke}, \citenamefont
  {Kreim}, \citenamefont {Mooser}, \citenamefont {Quint}, \citenamefont
  {Ulmer},\ and\ \citenamefont {Walz}}]{MainzProtonMagneticMoment}%
  \BibitemOpen
  \bibfield  {author} {\bibinfo {author} {\bibfnamefont {C.~C.}\ \bibnamefont
  {Rodegheri}}, \bibinfo {author} {\bibfnamefont {K.}~\bibnamefont {Blaum}},
  \bibinfo {author} {\bibfnamefont {H.}~\bibnamefont {Kracke}}, \bibinfo
  {author} {\bibfnamefont {S.}~\bibnamefont {Kreim}}, \bibinfo {author}
  {\bibfnamefont {A.}~\bibnamefont {Mooser}}, \bibinfo {author} {\bibfnamefont
  {W.}~\bibnamefont {Quint}}, \bibinfo {author} {\bibfnamefont
  {S.}~\bibnamefont {Ulmer}}, \ and\ \bibinfo {author} {\bibfnamefont
  {J.}~\bibnamefont {Walz}},\ }\href@noop {} {\bibfield  {journal} {\bibinfo
  {journal} {New Journal of Physics}\ }\textbf {\bibinfo {volume} {14}},\
  \bibinfo {pages} {063011} (\bibinfo {year} {2012})}\BibitemShut {NoStop}%
\end{thebibliography}

%

\end{document}